\documentclass[a4paper,12pt]{article}
\usepackage{amsmath,amsthm,amssymb,bbm}

\usepackage[english]{babel}

\title{\bf Entanglement robustness and geometry in systems of identical particles}

\author{F. Benatti$^{a,b}$, 
R. Floreanini$^{b}$, U. Marzolino$^{a,c}$\\
\\
\small ${}^a$Dipartimento di Fisica, Universit\`a di Trieste, 
34151 Trieste, Italy\\
\small ${}^b$Istituto Nazionale di Fisica Nucleare, Sezione di Trieste,
34151 Trieste, Italy\\
\small ${}^c$Physikalisches Institut, Albert-Ludwigs-Universitat
Freiburg, 79104 Freiburg, Germany}

\date{\null}

\begin{document}

\maketitle

\begin{abstract}
\noindent
The robustness properties of bipartite entanglement in systems of $N$
bosons distributed in $M$ different modes are analyzed using a definition of separability 
based on commuting algebras of observables, a natural choice
when dealing with identical particles. Within this framework,
expressions for the robustness and generalized robustness of entanglement can be 
explicitly given for large classes of boson states: their
entanglement content results in general much more stable than that of
distinguishable particles states. Using these results,
the geometrical structure of the space of $N$ boson states can be explicitly addressed.
\end{abstract}

\section{Introduction}

When dealing with many-body systems made of identical particles, the usual definitions
of separability and entanglement appear problematic since the natural particle
tensor product structure on which these notions are based is no longer available.
This comes from the fact that in such systems the microscopic constituents
can not be singly addressed, nor their properties directly measured
\cite{Feynman,Sakurai}.

This observation points to the need of generalized notions of separability and entanglement
not explicitly referring to the set of system states, or more in general 
to the ``particle'' aspect of first quantization: they should rather be based 
on the second quantized description of many-body systems,
in terms of the algebra of observables 
and to the behavior of the associated correlation functions.
This ``dual'' point of view stems from the fact that 
in systems of identical particles there is not a preferred 
notion of separability, it can be meaningful only in reference to a
given choice of (commuting) sets of observables \cite{Zanardi1}-\cite{Viola2}.

This new approach to separability and entanglement 
have been formalized in \cite{Benatti1}-\cite{Benatti3}:
it is valid in all situations, while reducing to the standard one for
systems of distinguishable particles;%
\footnote{The notion of entanglement
in many-body systems has been widely discussed 
in the recent literature ({\it e.g.}, see \cite{Schliemann}-\cite{Buchleitner});
nevertheless, we stress that
only a limited part of those results are relevant for identical particles systems.}
in particular, it has been successfully applied
to the treatment of the behavior of trapped ultracold bosonic gases,
giving rise to new, testable prediction in quantum metrology
\cite{Benatti1}-\cite{Argentieri}.

As in the case of systems of distinguishable particles,%
\footnote{For general reviews on the role of quantum correlations in systems
with large number of constituents see \cite{Lewenstein}-\cite{Amico}.}
suitable criteria able to detect non-classical correlations through
the implementations of practical tests are needed in order
to easily identify entangled bosonic many-body states \cite{Horodecki}-\cite{Modi}.
In the case of bipartite entanglement, the operation of partial transposition \cite{Peres}-\cite{Werner}
turns out to be again very useful; actually,
it has been found that in general this operation gives rise to a much more exhaustive
criterion for detecting bipartite entanglement than in the case of 
distinguishable particles \cite{Argentieri,Benatti3}. 
As a byproduct, this allows a rather complete classification
of the structure of bipartite entangled states in systems composed of $N$
bosons that can occupy $M$ different modes \cite{Benatti3},
becoming completely exhaustive in some relevant special cases, as for $M=2$.

In the following we shall further explore the properties of bipartite entanglement
in bosonic systems made of a fixed number of elementary constituents. We shall first
study to what extent an entangled bosonic state results robust against its
mixing with another state (separable or not): we shall find that in general, bosonic
entanglement is much more robust than the one of distinguishable particles.
In particular, we shall give an explicit expression for the so-called
``robustness'' \cite{Vidal} and upper bounds for the ``generalized robustness'' \cite{Steiner}.
As a byproduct, a characterization of the geometry of the space of
bosonic states will also be given; the structure
of this space results much richer than in the case of systems
with $N$ distinguishable constituents. One of the most striking results is that
the totally mixed separable state, proportional to the unit matrix, does no longer lay
in the interior of the subspace of separable states: in any of its neighborhood
entangled states can always be found.

\section{Entanglement of multimode boson systems}

As mentioned above, we shall focus on bosonic many-body systems made of $N$
elementary constituents that can occupy $M$ different modes. From the physical point of view,
this is a quite general framework, relevant in the study of quite different systems
in quantum optics, atom and condensed matter physics. For instance,
this theoretical paradigm is of special importance in modelling the behavior of ultracold bosonic gases
confined in multi-site optical lattices, that are becoming so relevant in the study of
quantum phase transitions and other quantum many-body effects ({\it e.g.}, see
\cite{Lewenstein,Bloch},
\cite{Stringari}-\cite{Yukalov} and references therein).

In order to properly describe the $N$ boson system, 
let us thus introduce creation $a^\dagger_i$ and annihilation 
operators $a_i$, $i=1, 2,\ldots,M$, for the $M$ different modes that the bosons can occupy,
obeying the standard canonical
commutation relations, $[a_i,\,a^\dagger_j]=\delta_{ij}$.
The total Hilbert space $\cal H$ of the system
is then spanned by the many-body Fock states, obtained by applying creation operators to the
vacuum:
\begin{equation}
|n_1, n_2,\ldots,n_M\rangle= {1\over \sqrt{n_1!\, n_2!\cdots n_M!}}
(a_1^\dagger)^{n_1}\, (a_2^\dagger)^{n_2}\, \cdots\, (a_M^\dagger)^{n_M}\,|0\rangle\ ,
\label{2-1}
\end{equation}
the integers $n_1, n_2, \ldots, n_M$ representing the occupation numbers of the different modes.
Since the number of bosons is fixed,
the total number operator $\sum_{i=1}^M a_i^\dagger a_i$
is a conserved quantity and the occupation numbers must satisfy the 
additional constraint $\sum_{i=1}^M n_i=N$; in other words, all states must
contain exactly $N$ particles. As a consequence, the dimension $D$ of the system Hilbert space $\cal H$
is finite; one easily finds: $D={N+M-1\choose N}$.

In addition, the set of polynomials in all creation and annihilation operators,
$\{a^\dagger_i,\, a_i\}$, $i=1,2,\ldots, M$,
form an algebra that, together with its norm-closure, coincides with the algebra
${\cal B}({\cal H})$ of bounded operators;%
\footnote{The algebra ${\cal B}({\cal H})$ is generated by the so-called Weyl operators;
all polynomials in the in the creation and annihilation operators are obtained from them
by proper differentiation \cite{Thirring,Strocchi}.}
the observables of the systems are part of this algebra.

When dealing with systems of identical particles, 
instead of focusing on partitions of the Hilbert space $\cal H$, it seems natural to define the notion
of bipartite entanglement by the presence of non-classical correlations among 
observables belonging to two commuting subalgebras ${\cal A}_1$ and 
${\cal A}_2$ of ${\cal B}({\cal H})$ \cite{Benatti1}.
Quite in general, one can then introduce the following definition:

\medskip
\noindent
{\bf Definition 1.} {\sl An {\bf algebraic bipartition} of the algebra ${\cal B}({\cal H})$ is any pair
$({\cal A}_1, {\cal A}_2)$ of commuting subalgebras of ${\cal B}({\cal H})$,
${\cal A}_1, {\cal A}_2\subset {\cal B}({\cal H})$.}
\medskip

\noindent
More explicitly, a bipartition
of the $M$-oscillator algebra ${\cal B}({\cal H})$ can be given by splitting the collection
of creation and annihilation operators into two disjoint sets,
$\{a_i^\dagger,\, a_i\, | i=1,2\ldots,m\}$ and 
$\{a_j^\dagger,\, a_j,\, |\, j=m+1,m+2,\ldots,M\}$; 
it is thus uniquely determined by
the choice of the integer $m$, with $0\leq m \leq M$.
All polynomials in the first set (together with their norm-closures)
form a subalgebra ${\cal A}_1$, while the remaining set analogously generates
a subalgebra ${\cal A}_2$.
Since operators pertaining to different modes
commute, one sees that any element of the subalgebra ${\cal A}_1$ 
commutes with each element of ${\cal A}_2$, 
in short $[{\cal A}_1, {\cal A}_2]=\,0$.

\smallskip

\noindent
{\bf Remark 1:} {\sl i)} Note that there is no loss of generality 
in assuming the modes forming the subalgebras ${\cal A}_1$
(and ${\cal A}_2$) to be contiguous; if in the chosen bipartition this is not the case, 
one can always re-label the modes in such a way to achieve this convenient ordering.\hfill\break
{\sl ii)} Further, when the two commuting algebras ${\cal A}_1$ and ${\cal A}_2$ are generated
only by a subset $M'<M$ of modes, one can simply proceed 
as if the $N$ boson system would contain just the used $M'$ modes, since
all operators in ${\cal B}({\cal H})$ pertaining to the modes not involved in the bipartition
commute with any element of the two subalgebras ${\cal A}_1$ and ${\cal A}_2$,
and therefore effectively act as ``spectators''. As a consequence, all the considerations
and results discussed below holds also in this situation, provided one replaces
the total number of modes $M$ with $M'$, the actual number of modes used 
in the chosen bipartition.\hfill$\Box$
\smallskip

Having introduced the notion of algebraic bipartition $({\cal A}_1, {\cal A}_2)$
of the system operator algebra ${\cal B}({\cal H})$, one can now define
the notion of local observable:

\medskip
\noindent
{\bf Definition 2.} {\sl An element (operator) of ${\cal B}({\cal H})$ is said to be 
$({\cal A}_1, {\cal A}_2)$-{\bf local}, {\it i.e.} local with respect to
a given bipartition $({\cal A}_1, {\cal A}_2)$, if it is the product $A_1 A_2$ of an element 
$A_1$ of ${\cal A}_1$ and another $A_2$ in ${\cal A}_2$.}

\medskip
\noindent
From this notion of operator locality, a natural definition of state separability and entanglement
follows \cite{Benatti1}:

\medskip
\noindent
{\bf Definition 3.} {\sl A state $\rho$ (density matrix) will be called {\bf separable} with
respect to the bipartition $({\cal A}_1, {\cal A}_2)$, in short $({\cal A}_1, {\cal A}_2)$-separable,
if the expectation ${\rm Tr}\big[\rho\, A_1 A_2\big]$ 
of any local operator $A_1 A_2$ can be decomposed into a linear convex combination of
products of expectations:
\begin{equation}
{\rm Tr}\big[\rho\, A_1 A_2\big]=\sum_k\lambda_k\, {\rm Tr}\big[\rho_k^{(1)}\, A_1\big]\, 
{\rm Tr}\big[\rho_k^{(2)}\, A_2\big]\ ,\qquad
\lambda_k\geq0\ ,\quad \sum_k\lambda_k=1\ ,
\label{2-2}
\end{equation}
where $\{\rho_k^{(1)}\}$ and $\{\rho_k^{(2)}\}$ are collections of admissible states for the systems;
otherwise the state $\rho$ is said to be {\bf entangled} with respect the bipartition
$({\cal A}_1, {\cal A}_2)$.}
\medskip

\noindent
{\bf Remark 2:} {\sl i)} This generalized definition of separability can easily be extended to the case
of more than two partitions; for instance, in the case of an $n$-partition, 
Eq.(\ref{2-2}) above would extend to:
\begin{equation}
{\rm Tr}\big[\rho\,A_1 A_2\cdots A_n]=\sum_k\lambda_k\, {\rm Tr}\big[\rho_k^{(1)}A_1\big]\, 
{\rm Tr}\big[\rho_k^{(2)}A_2\big]\cdots
{\rm Tr}\big[\rho_k^{(n)}A_n\big]\, ,\quad
\lambda_k\geq0\ ,\,\, \sum_k\lambda_k=1\ .
\label{2-3}
\end{equation}

\noindent
{\sl ii)} When dealing with systems of {\sl distinguishable} particles, one finds
that {\sl Definition 3} gives the standard notion of separability \cite{Benatti1}.\hfill\break
\noindent
{\sl iii)} In this respect, it should be noticed that when dealing with systems of identical particles, 
there is no {\it a priori} given, natural partition,
so that questions about entanglement and separability, non-locality and locality 
are meaningful only with reference to a specific choice 
of commuting algebraic sets of observables \cite{Zanardi1}-\cite{Benatti3}; 
this general observation, often overlooked in the literature, is at the basis of the definitions
given in (\ref{2-2}) and (\ref{2-3}).\hfill\break
\noindent
{\sl iv)} A special situations is represented by pure states \cite{Benatti3}. 
In fact, when dealing with pure states,
instead of general statistical mixtures, and bipartitions that involve the whole algebra
${\cal B}({\cal H})$, the separability condition in (\ref{2-2}) (and similarly for (\ref{2-3}))
simplify, becoming:
\begin{equation}
{\rm Tr}\big[\rho\, A_1 A_2\big]={\rm Tr}\big[\rho^{(1)}\, A_1\big]\, 
{\rm Tr}\big[\rho^{(2)}\, A_2\big]\ ,\quad\rho=|\psi\rangle\langle\psi|\ ,
\label{2-4}
\end{equation}
with $\rho^{(1)}$, $\rho^{(2)}$ projections on the restrictions of $|\psi\rangle$ 
to the first, respectively second, partition;
in other terms, separable, pure states turn out to be just product states. \hfill$\Box$

\smallskip

Examples of $N$ bosons pure separable states are the Fock states.
Using the notation and specifications introduced before ({\it cf.} (\ref{2-1})),
they can be recast in the form:
\begin{equation}
| k_1, \ldots, k_m\,;\, k_{m+1}, \ldots, k_M\rangle\ ,\quad
\sum_{i=1}^m k_i =k\ ,\ \sum_{j=m+1}^M k_j=N-k\ ,\ \ 0\leq k \leq N\ ,
\label{2-5}
\end{equation}
where $k$ indicates the number of bosons in the first partition; by varying it together with
the integers $k_i$, these states generate the whole Hilbert space $\cal H$.
This basis states can be relabeled in a different, more convenient way as:
\begin{equation}
| k, \sigma; N-k, \sigma'\rangle\ ,\quad \sigma=1,2, \ldots, {k+m-1\choose k}\ ,\
\sigma'=1, 2,\ldots, {N-k+M-m-1\choose N-k}\ ,
\label{2-6}
\end{equation}
where, as before, the integer $k$ represents the number of particles found in the first $m$ modes,
while $\sigma$ counts the different ways in which those particles can fill those modes;
similarly, $\sigma'$ labels the ways in which the remaining $N-k$ particles
can occupy the other $M-m$ modes.%
\footnote{Clearly, we need two extra labels $\sigma$ and $\sigma'$ for each value of $k$, so that
these labels (as well the range of values they take) are in general $k$-dependent: in order to keep
the notation as a simple as possible, in the following these dependences will be tacitly understood.}
In this new labelling, the property of orthonormality of the states
in (\ref{2-5}) becomes:
$\langle k, \sigma; N-k, \sigma'| l, \tau; N-l, \tau'\rangle=
\delta_{kl}\,\delta_{\sigma\tau}\,\delta_{\sigma'\tau'}$.

For fixed $k$, the basis vectors $\{| k, \sigma; N-k, \sigma'\rangle\}$ span a subspace ${\cal H}_k$
of dimension $D_k\, D_{N-k}$, where for later convenience we have defined ({\it cf.} (\ref{2-6}) above):
\begin{equation}
D_k\equiv{k+m-1\choose k}\ ,\qquad
 D_{N-k}\equiv{N-k+M-m-1\choose N-k}\ .
\label{2-7}
\end{equation}

\noindent
{\bf Remark 3:} Note that the space ${\cal H}_k$ is naturally isomorphic to the tensor
product space $\mathbb{C}^{D_k}\otimes \mathbb{C}^{D_{N-k}}$; through this isomorphism, the states
$| k, \sigma; N-k, \sigma'\rangle$ can then be identified with the
corresponding basis states of the form $| k, \sigma\rangle \otimes | N-k, \sigma'\rangle$.
\hfill$\Box$
\medskip

\noindent
By summing over all values of $k$, thanks to a known binomial summation formula \cite{Prudnikov},
one recovers the dimension $D$ of the whole Hilbert space $\cal H$
:
\begin{equation}
\sum_{k=0}^N D_k\, D_{N-k}=D= {N+M-1\choose N}\ .
\label{2-8}
\end{equation}
Using this notation, a generic mixed state $\rho$
can then be written as:
\begin{equation}
\rho=\sum_{k,l=0}^N\ \sum_{\sigma,\sigma',\tau,\tau'}\
\rho_{k \sigma\sigma', l\tau\tau'}\ | k, \sigma; N-k, \sigma'\rangle \langle l, \tau; N-l, \tau' |\ ,
\quad \sum_{k=0}^N\ \sum_{\sigma,\sigma'}\
\rho_{k \sigma\sigma', k\sigma\sigma'}=1\ .
\label{2-9}
\end{equation}

In general, to determine whether a given density matrix $\rho$ can be written in separable 
form is a hard task and one is forced to rely on suitable separability tests.
One of the most useful such criteria involves the operation
of partial transposition \cite{Peres,Horodecki2}: 
a state $\rho$ for which the partially transposed density matrix $\tilde\rho$ is
no longer positive is surely entangled. This lack of positivity can be witnessed by the
so-called negativity \cite{Zyczkowski,Vidal1,Horodecki}:
\begin{equation}
{\cal N}(\rho)=\, {1\over2}\Big(\!||\tilde\rho||_1 - {\rm Tr}[\rho]\Big)\ ,\qquad 
||\tilde\rho||_1={\rm Tr}\Big[\sqrt{\tilde\rho^\dagger \tilde\rho}\Big]\ .
\label{2-10}
\end{equation}
which is nonvanishing only in presence of a non positive $\tilde\rho$.
Although this criterion is not exhaustive (there are entangled states that remain
positive under partial transposition), it results much more reliable in systems
made of identical particles \cite{Argentieri,Benatti3}. 
Indeed, the operation of partial transposition 
gives a necessary and sufficient criteria for entanglement detection for very general
classes of bosonic states (\ref{2-9}), {\it e.g.} in presence of only two modes ($M=2$), or, in the
generic case of arbitrary
$M$, when the $({\cal A}_1,\, {\cal A}_2)$-bipartition is such that the algebra
${\cal A}_1$ is generated by creation and annihilation operators of just one mode,
while the remaining $M-1$ modes generates ${\cal A}_2$.

Even more interestingly, it turns out that entangled $N$-body bosonic states need to be 
of a definite, specific form \cite{Benatti3}:

{\bf Proposition 1.} {\sl A generic $(m, M-m)$-mode bipartite state (\ref{2-9}) is entangled
if and only if it can not be cast in the following block diagonal form
\begin{equation}
\rho=\sum_{k=0}^N p_k\ \rho_k\ ,\qquad \sum_{k=0}^N p_k=1\ ,\quad {\rm Tr}[\rho_k]=1\ ,
\label{2-11}
\end{equation}
with
\begin{equation}
\rho_k=\sum_{\sigma,\sigma',\tau,\tau'}\
\rho_{k \sigma\sigma', k\tau\tau'}\ | k, \sigma; N-k, \sigma'\rangle \langle k, \tau; N-k, \tau' |\ ,
\quad \sum_{\sigma,\sigma'}\rho_{k \sigma\sigma', k\sigma\sigma'}=1\ ,
\label{2-12}
\end{equation}
({\it i.e.} at least one of its non-diagonal coefficients $\rho_{k \sigma\sigma', l\tau\tau'}$, $k\neq l$,
is nonvanishing),
or if it can, at least one of its diagonal blocks $\rho_k$ is non-separable.}%
\footnote{For each block $\rho_k$, separability is understood with reference to
the isomorphic structure $\mathbb{C}^{D_k}\otimes \mathbb{C}^{D_{N-k}}$ 
mentioned before (see {\sl Remark 3}).}
\medskip

\noindent
{\sl Proof.} Assume first that the state $\rho$ can not be written in block diagonal form;
one can then show \cite{Benatti3} that it is not left positive by the operation
of partial transposition and therefore it is entangled. Next, take $\rho$
in block diagonal form as in (\ref{2-11}), (\ref{2-12}) above. If all its blocks
$\rho_k$ are separable, then clearly $\rho$ itself results separable.
Then, assume that at least one of the diagonal blocks
is entangled. By mixing it with the remaining blocks as in (\ref{2-11}) will not
spoil its entanglement since all blocks $\rho_k$ have support on orthogonal spaces;
as a consequence, the state $\rho$ results itself non-separable.
\hfill$\Box$

\medskip
Having found the general form of non-separable $N$-boson states, one can next ask
how robust is their entanglement content against mixture with other states.
This question has been extensively studied for states of distinguishable particles 
\cite{Vidal,Steiner,Vidal1,Plenio,Horodecki};
in the next section we shall analyze to what extent the results obtained in that case
can be extended to systems with a fixed number of bosons.

\section{Robustness of entanglement}

Several measures of entanglement have been introduced in the literature with the aim of 
characterizing quantum correlations and its usefulness in specific applications
\cite{Horodecki}-\cite{Modi}.
Starting with the entanglement of formation, most of these measures point to the
quantification of the entanglement content of a given state. A different approach
to this general problem has been proposed in \cite{Vidal,Steiner}: the idea is to obtain information
about the magnitude of non-classical correlations contained in 
a state $\rho$ by studying how much it can be mixed with other states before
becoming separable. 

More precisely, let us indicate with $\cal M$ the set of all
systems states and by ${\cal S}\subset{\cal M}$ that of separable ones;
then, with reference to an arbitrary $(m, M-m)$-mode bipartition, one can introduce the
following definition:

\medskip
\noindent
{\bf Definition 4.} {\sl Given a state $\rho$, its {\bf robustness of entanglement} 
is defined by
\begin{equation}
R(\rho)={\rm inf}\Big\{ t\ |\ t\geq0,\ \exists\, \sigma\in{\cal M}'\subset{\cal M},\ {\rm for\ which} \
\eta\equiv{\rho + t\,\sigma\over(1+t)} \in{\cal S}\Big\}\ ,
\label{3-1}
\end{equation}
{\it i.e.} it is the smallest, non-negative $t$ such that a state $\sigma$ exists for which the
(unnormalized) combination $\rho + t\,\sigma$ is separable.}
\medskip

\noindent
Actually, various forms of robustness have been introduced: they all share the definition
(\ref{3-1}), but differ in the choice of the subset
${\cal M}'$ from which the mixing state $\sigma$ should be drawn.
In particular, one talks of {\it generalized robustness} $R_g$ when $\sigma$ can be any state \cite{Steiner},
while simply of {\it robustness} $R_s$ when $\sigma$ must be separable \cite{Vidal}.

All robustness defined in this way satisfy nice properties; more specifically, they result
entanglement monotones, {\it i.e.} they are invariant under local operations and classical communication,
and convex, $R\big(\lambda\, \rho_1+(1-\lambda)\rho_2\big)\leq \lambda\, R(\rho_1) +
(1-\lambda)\, R(\rho_2)$; further, $R(\rho)$ in (\ref{3-1}) is vanishing if and only if
$\rho$ itself is separable. Although implicitly proven in the case of states of distinguishable
particles \cite{Vidal,Steiner}, these properties hold true also in the case of $N$-boson systems. Nevertheless,
there are striking differences in the behavior of the robustness of states of identical particles
with respect to what is known in systems with distinguishable constituents.

Let us first focus on the robustness $R_s(\rho)$, that measures how strong is the entanglement
content of a state $\rho$ when mixed with separable states. One finds:

\medskip
\noindent
{\bf Proposition 2.} {\sl The robustness of entanglement of a generic $(m, M-m)$-mode
bipartite state $\rho$ is given by
\begin{equation}
R_s(\rho)=\sum_{k=0}^N p_k\, R_s(\rho_k)\ ,
\label{3-2}
\end{equation}
for states that are in block diagonal form as in (\ref{2-11}), (\ref{2-12}), while it is infinitely large
otherwise.}
\medskip

\noindent
{\sl Proof.} From the results of {\it Proposition 1}, we know that separable $N$-boson states
must be block diagonal. If the state $\rho$ is not in this form, it can
never be made block diagonal by mixing it with any separable one; therefore, in this case, the combination 
$\rho + t\, \sigma$ will never be separable, unless $t$ is infinitely large.

Next, consider the case in which the state $\rho$ is in block diagonal
form, {\it i.e.} it can be written as in (\ref{2-11}), (\ref{2-12}). 
First, if $\rho$ is separable, then clearly $R_s(\rho)=\,0$.
For an entangled $\rho$, one can discuss each block $\rho_k$ separately: this is allowed 
since they have support on orthogonal Hilbert subspaces. Then, let us indicate by $t_k$ the robustness
of block density matrix $\rho_k$; by {\sl Remark 3} and the definition of robustness, 
the numbers $t_k$'s are finite and positive,
vanishing only when the corresponding state $\rho_k$ is separable \cite{Vidal}. More specifically, for each
$k$, there exist separable states $\sigma_k$ and $\eta_k$, such that:
\begin{equation}
\rho_k +t_k\, \sigma_k =(1+t_k)\, \eta_k\ .
\label{3-3}
\end{equation}
Multiplying both sides of this relation by the positive number $p_k$ and then summing over $k$,
one gets
\begin{equation}
\rho +t\, \sigma =(1+t)\, \eta\ ,\qquad t=\sum_{k=0}^N p_k\, t_k\ ,
\label{3-4}
\end{equation}
where the separable states $\sigma$ and $\eta$ are explicitly given by
\begin{equation}
\sigma=\sum_{k=0}^N \bigg({p_k\, t_k\over t}\bigg)\ \sigma_k\ ,\qquad
\eta=\sum_{k=0}^N \bigg({1+t_k \over 1+t}\bigg)\, p_k\, \eta_k\ .
\label{3-5}
\end{equation}
To prove that indeed $t$ given in (\ref{3-4}) is really the robustness of $\rho$, one needs to 
check that no better decomposition 
\begin{equation}
\rho +t'\, \sigma' =(1+t')\, \eta'\label{}\ ,
\label{3-6}
\end{equation}
with $t'\leq t$, exists. In order to show this, let proceed {\it ad absurdum} 
and assume that such decomposition can indeed be found.
Since the states $\sigma'$ and $\eta'$ are separable, they must be block diagonal, {\it i.e.}
of the form $\sigma'=\sum_k q_k\, \sigma'_k$ and $\eta'=\sum_k r_k\, \eta'_k$
with $\sum_k q_k=\sum_k r_k=1$ and $\sigma'_k$ and $\eta'_k$ separable density matrices.
By the orthogonality of the Hilbert subspaces with fixed $k$, from (\ref{3-6}) one then gets
\begin{equation}
p_k\, \rho_k + t'\, q_k\, \sigma'_k = (1+t')\, r_k\, \eta'_k\ ,
\label{3-7}
\end{equation}
and further, by taking its trace, $p_k+ t'\, q_k = (1+t')\, r_k$. In addition, from the previous identity,
one sees that the combination
\begin{equation}
\rho_k + t'\, {q_k\over p_k}\, \sigma'_k\ ,
\label{3-8}
\end{equation}
is separable. By definition of robustness of the block $\rho_k$ as given in (\ref{3-3}),
it then follows that:
\begin{equation}
t'\, {q_k\over p_k}\geq t_k\ ,
\label{3-9}
\end{equation}
ore equivalently, $ t_k\, p_k \leq t'\, q_k$. By summing over $k$, one then finds:
$t\leq t'$; this result is compatible with the initial assumption $t\geq t'$ 
only if $t'$ coincides with $t$. Therefore, the robustness of the block diagonal state $\rho$
is indeed given by the weighted sum of the robustness of each block. \hfill$\Box$

\medskip
The problem of finding the robustness $R_s(\rho)$ of a generic $N$-boson state $\rho$ is then
reduced to the more manageable task of identifying the robustness of its diagonal
blocks, which are finite-dimensional density matrices 
for which standard techniques and results can be used.
\medskip

\noindent
{\bf Remark 4:} {\sl i)} A remarkable property of the robustness of entanglement of states describing
distinguishable particles is that it is equal to the negativity for pure states. 
In the case of identical particles, this property does not hold anymore as the robustness of entanglement 
of non-block diagonal pure states is infinitely large. Nevertheless, one can easily show that in
general for pure states: ${\cal N}(\rho)\leq R_s(\rho)$. \hfill\break
{\sl ii)} The robustness of entanglement of states that, with respect to the given partition, result
mixtures of pure block states, $\rho=\sum_k p_k\rho_k$, $\rho_k{}^2=\rho_k$, is equal 
to their negativity $R_s(\rho)=\sum_k p_k\, {\cal N}(\rho_k)={\cal N}(\rho)$,
since now $R_s(\rho_k)={\cal N}(\rho_k)$, as in the standard case (see \cite{Benatti3}). \hfill $\Box$
\medskip

More difficult appears the task of computing the generalized robustness
$R_g(\rho)$ of a generic $N$-boson state: only upper bounds can in general be given.
In any case, note that in general $R_g(\rho)\leq R_s(\rho)$ since the optimization procedure of
{\sl Definition 4} is performed over a larger subset of states in the case of the
generalized robustness.

By fixing as before a $(m, M-m)$-mode bipartition, a first bound on $R_g(\rho)$ 
can be easily obtained. Let us extract from $\rho$ its 
diagonal part $\rho_D$, as defined in terms of 
a Fock basis determined by the given bipartition ({\it cf.} (\ref{2-6})),
and call $\rho_{ND}\equiv \rho-\rho_D$ the rest. By definition of separability,
$\rho_D$ is surely separable with respect to 
the chosen bipartition, so that an easy way to get a separable state
by mixing $\rho$ with another state $\sigma$ is to subtract from it its non-diagonal part.
However, $-\rho_{ND}$ alone is not in general a density matrix since it might have negative
eigenvalues. Let us denote by $\lambda$ the modulus of its largest negative eigenvalue;
then the quantity $\lambda \mathbbm{1} -\rho_{ND}$, where $\mathbbm{1}$ is the identity matrix,
will surely be positive and therefore, once
normalized, can play the role of the density matrix $\sigma$ in the separable combination
$\rho +t\, \sigma\equiv\rho_D +\lambda\, \mathbbm{1}$. By taking the trace, one finds
for the normalization factor $t$ the following expression: $t=\lambda\, D$,
where $D$ is as before the dimension of the total Hilbert space. By definition of robustness,
it then follows that 
\begin{equation}
R_g(\rho)\leq \lambda\, D
\label{3-10}\ ;
\end{equation}
as a consequence, the generalized robustness
of a generic $N$-boson state is always finite.

A different bound on $R_g(\rho)$ can be obtained using a refined decomposition for $\rho$,
\begin{equation}
\rho=\rho_B + \rho_{NB}\ ,\qquad 
\rho_B=\sum_{k=0}^N p_k\ \rho_k\ ,\qquad \sum_{k=0}^N p_k=1\ ,\quad {\rm Tr}[\rho_k]=1\ ,
\label{3-11}
\end{equation}
where $\rho_B$ is the block diagonal part of $\rho$, whose blocks $\rho_k$ can be written
as in (\ref{2-12}), while $\rho_{NB}\equiv\rho-\rho_B$ is the rest, containing the
non block diagonal pieces. One can first ask for the generalized robustness
of $\rho_B$, which is a {\it bona fide} state, being a normalized, positive
matrix.%
\footnote{This is a direct consequence of the positivity of $\rho$, since
$\rho_B$ is made of its principal minors.}
Quite in general, one has:

\medskip
\noindent
{\bf Proposition 3.} {\sl The generalized robustness of entanglement of a generic $(m, M-m)$-mode
bipartite state $\rho$ given in block diagonal form as in (\ref{2-11}), (\ref{2-12}) is given by
\begin{equation}
R_g(\rho)=\sum_{k=0}^N p_k\, R_g(\rho_k)\ .
\label{3-12}
\end{equation}
}
\medskip

\noindent
{\sl Proof.} It is the same as in {\sl Proposition 2\,}; the only difference
is that now the states $\sigma_k$ are in general entangled. \hfill $\Box$
\medskip

\noindent
As a consequence, for a generic state as in (\ref{3-11}) above, due to the presence of the
additional term $\rho_{NB}$, one surely has:
$R_g(\rho)\geq R_g(\rho_B)=\sum_{k=0}^N p_k\, R_g(\rho_k)$.

To get an upper bound for $R_g(\rho)$, let us consider the form that the
(unnormalized) density matrices $\sigma$ must take in order to make 
the combination $\rho+\sigma$ separable; by \hbox{\sl Definition 4}, the generalized
robustness of entanglement coincides with minimum value of their traces:
$R_g(\rho)={\rm inf}\big\{{\rm Tr}[\sigma]\big\}$.
Since the combination $\rho+\sigma$ is separable, it must be in block diagonal form;
therefore, $\sigma$ must surely contain the contribution $-\rho_{NB}$. Further,
it must also take care of the entanglement in the remaining block
diagonal term $\rho_B$; in view of the results of {\sl Proposition 3},
this can be obtained (and in an optimal way) 
by the contribution $\sum_{k=0}^N p_k\, R_g(\rho_k)\, \sigma_k$,
where $\sigma_k$ is the optimal density matrix that makes the diagonal block $\rho_k$
separable.

However, while $\sum_{k=0}^N p_k\, R_g(\rho_k)\, \sigma_k$ is a positive matrix, in general $\rho_{NB}$
is not; therefore $\sigma$ should contain a further contribution, 
a (unnormalized) positive and separable matrix $\tilde\sigma$ curing the negativity
induced by $\rho_{NB}$. As a consequence, the generic form of the positive matrix $\sigma$
making the combination $\rho+\sigma$ separable is given by:
\begin{equation}
\sigma=\sum_{k=0}^N p_k\, R_g(\rho_k)\, \sigma_k -\rho_{NB} + \tilde\sigma\ ,
\label{3-14}
\end{equation}
and the computation of the generalized robustness of $\rho$ is reduced to the 
determination of the optimal $\tilde\sigma$; indeed:
\begin{equation}
R_g(\rho)=\sum_{k=0}^N p_k\, R_g(\rho_k) + {\rm inf}\big\{{\rm Tr}[\tilde\sigma]\big\}\ .
\label{3-15}
\end{equation}
Upper bounds on $R_g(\rho)$ can then be obtained by estimating the above minimum value,
trough specific choices of $\tilde\sigma$.

A simple possibility for curing the non positivity of $-\rho_{NB}$ is to add to it
a matrix proportional to the modulus of its largest negative eigenvalue; 
in general, the value of this eigenvalue is however
difficult to estimate. Another possibility is suggested by the general theory
of positive matrices (see Theorem 6.1.1 and 6.1.10 in \cite{Horn}): 
a sufficient condition for a generic hermitian matrix $M_{ij}$ to be positive is that it must be
``diagonally dominated'', {\it i.e.} $M_{ii}\geq \sum_{j\neq i} |M_{ij}|$, $\forall i$.%
\footnote{Note that this condition is base-dependent: nonequivalent conditions
are obtained by expressing the matrix $M$ in different basis.}
Then, in a fixed separable basis, by choosing for $\tilde\sigma$
the diagonal matrix whose entries are given by
the sum of the modulus of the elements of the corresponding rows of $-\rho_{NB}$,
the matrix $\sigma$ in (\ref{3-14}) results positive and
makes the combination $\rho+\sigma$ separable. One can then conclude that:
\begin{equation}
R_g(\rho)\leq\sum_{k=0}^N p_k\, R_g(\rho_k) + ||\rho_{NB}||_{\ell_1}\ ,
\label{3-16}
\end{equation}
where, for any matrix $M$, $|| M||_{\ell_1}=\sum_{i,j} |M_{ij}|$ is the so-called
$\ell_1$-norm (see \cite{Horn}).%
\footnote{The same procedure can also be applied to the previously used decomposition
of $\rho$ into its diagonal and off-diagonal parts: $\rho=\rho_D +\rho_{ND}$;
the mixing matrix $\sigma$ would now be composed by $-\rho_{ND}$ plus a diagonal matrix
whose entries are given by the sums of the modulus of the elements of each row of $\rho_{ND}$.
In this case, one easily finds that: $R_g(\rho)\leq ||\rho_{ND}||_{\ell_1}$;
although in general $||\rho_{ND}||_{\ell_1} \geq ||\rho_{NB}||_{\ell_1}$, this constitutes a
different upper bound for the generalized robustness, independent from that given in (\ref{3-16}).}

In presence of just two modes, $M=2$, each of which forming a partition,
the above considerations further simplify. In this case,
the Fock basis in (\ref{2-6}) is given by the set of $N+1$ vectors $\{ |k;N-k\rangle,\ 0\leq k\leq N \}$,
without the need of further labels; indeed, the $N$ bosons can occupy either one
of the two modes. 
Notice that by (\ref{2-4}) this set of Fock vectors 
constitutes the only basis made of separable pure states \cite{Benatti3}. 
In this basis, a generic density matrix for the system can then be written as:
\begin{equation}
\rho=\sum_{k,l=0}^N\ \rho_{kl}\ | k; N-k\rangle \langle l; N-l |\ ,
\quad \sum_{k=0}^N\ \rho_{kk}=1\ .
\label{3-17}
\end{equation}
By {\sl Proposition 1}, once adapted to this simplified case, 
it follows that a state as in (\ref{3-17}) is separable 
if and only if $\rho_{kl}\sim\delta_{kl}$, {\it i.e.}
the density matrix $\rho$ is diagonal in the Fock basis. As a consequence,
an entangled state can never be made separable by mixing it with a separable state,
so that its robustness of entanglement results always infinite.

In the case of the generalized robustness, since there are only diagonal and off-diagonal
terms and no blocks, the above discussed upper bounds (\ref{3-10}) and (\ref{3-16}) simplify, becoming:
\begin{equation}
i)\ R_g(\rho)\leq \lambda\, (N+1)\ ,\qquad ii)\ 
R_g(\rho)\leq ||\rho_{ND}||_{\ell_1}\ ,
\label{3-18}
\end{equation}
where, as before, $\rho_{ND}$ is the non diagonal part of $\rho$ in the Fock basis, and
$\lambda$ the modulus of its largest positive eigenvalue. 
These bounds can be explicitly evaluated for specific classes of states as shown below.
\medskip

\noindent
{\bf Examples:} {\sl i)} Let us first consider pure states of the form:
\begin{equation}
\rho\equiv|\psi\rangle\langle\psi|\ ,\qquad
|\psi\rangle= {1\over\sqrt{N+1}}\sum_{k=0}^N p_k\, |k;N-k\rangle\ ,\qquad  p_k=e^{i\varphi_k}\ .
\label{3-19}
\end{equation}
The non-diagonal part of the matrix $\rho$ is given by $\rho_{ND}=(\Phi-\mathbbm{1})/(N+1)$
where $\Phi=\sum_{k,l} e^{i(\varphi_k-\varphi_l)}\, |k;N-k\rangle \langle l;N-l|$
is the $(N+1)\times(N+1)$ matrix of phase differences; its eigenvalues are
zero and $N+1$, so that the largest negative eigenvalue of $-\rho_{ND}$ is in modulus
$N/(N+1)$. From the first bound in (\ref{3-18}) one therefore gets: $\ R_g(\rho)\leq N$.
This is also the result of the second bound, since the norm $||\rho_{ND}||_{\ell_1}$
is also equal to $N$.\hfill\break
{\sl ii)} Nevertheless, the two bounds in (\ref{3-18}) 
do not in general coincide, as can be seen by slightly generalizing the
states in (\ref{3-19}) by allowing the coefficients $p_k$ to acquire a non unit norm, $p_k=|p_k| e^{i\varphi_k}$, 
$\sum_k |p_k|^2=1$. By choosing the norms and phases of the $p_k$'s to be uniformly distributed, one can
easily generate states for which the second bound in (\ref{3-18}) 
is more stringent.\hfill\break
{\sl iii)} Note, however, that the hierarchy of the bounds in (\ref{3-18}) can
be reversed, as it happens for instance with the following mixed state:
\begin{equation}
\rho=\frac{1}{N+1}\sum_{k=0}^N |k;N-k\rangle\langle k;N-k|-\frac{1}{N(N+1)}
\sum_{\substack{k,l=0\\k\neq l}}^N |k;N-k\rangle\langle l;N-l|\ .
\label{3-20}
\end{equation}
Indeed, now one has
$\rho_{ND}=(\mathbbm{1}-E)/N(N+1)$, where $E$ is the matrix with all entries equal to one.
One easily checks that $||\rho_{ND}||_{\ell_1}=1$, while the modulus 
of the largest negative eigenvalue $\lambda$
of $-\rho_{ND}$ is $1/N(N+1)$. Therefore, from the first bound in (\ref{3-18}) one gets
$R_g(\rho)\leq 1/N$, which is lower than the second one.\hfill $\Box$

\section{On the geometry of $N$-boson states}

As shown by the previous results, the properties of the states of a system of $N$-bosons
result rather different and to a certain extent richer
then those describing distinguishable particles. This is surely a consequence
of the indistinguishability of the system elementary constituents, but also
of the presence of the additional constraint that fixes the total number of
particles to be $N$. This additional ``rigidity'' allows nevertheless a detailed
description of the geometric structure of the set of $N$-boson states.

Let us first consider the set of entangled states. As mentioned before,
for systems of identical particles, negativity, as defined in (\ref{2-10}),
results a much more exhaustive entanglement criteria than in systems
of distinguishable particles: it seems then appropriate to look for states
that maximize it.

\medskip
\noindent
{\bf Proposition 4.} {\sl Given any bipartition $(m, M-m)$, 
the negativity ${\cal N}(\rho)$ is maximized by pure states
that have all the Schmidt coefficients nonvanishing and equal to a normalizing constant.}

\medskip

\noindent
{\sl Proof.} First observe that the negativity is a convex function \cite{Vidal1}, {\it i.e.} it satisfies the inequality
\begin{equation}
{\cal N}\Big(\sum_i p_i\,\rho_i\Big)\leq \sum_i p_i\, {\cal N}(\rho_i)\ ,\qquad p_i\geq0\ ,\qquad
\sum_i p_i=1\ ,
\label{4-1}
\end{equation}
for any convex combination $\sum_i p_i\,\rho_i$
of density matrices $\rho_i$. Since any density matrix can be written as a convex combination
of projectors over pure states, one can limit the search to pure states;
in a given $(m, M-m)$-bipartition, they can be expanded in terms of the Fock basis (\ref{2-6}) as:
\begin{equation}
|\psi\rangle=\sum_{k=0}^N |\psi_k\rangle\ ,\qquad
|\psi_k\rangle\equiv\sum_{\sigma=1}^{D_k}\sum_{\sigma'=1}^{D_{N-k}} 
\psi_{k\sigma\sigma'}\, |k,\sigma;N-k,\sigma'\rangle\ ,\qquad 
\sum_k\sum_{\sigma\sigma'} |\psi_{k\sigma\sigma'}|^2=1\ .
\label{4-2}
\end{equation}
As observed before, for each $k$, the set of vectors $\{|k,\sigma;N-k,\sigma'\rangle\}$
span a subspace ${\cal H}_k\subset {\cal H}$ of finite dimension 
$D_k\, D_{N-k}$ of the total Hilbert space $\cal H$, and the component $|\psi_k\rangle$ is a vector of this
space. Recalling {\sl Remark 3}, one can then write it in Schmidt form,
\begin{equation}
|\psi_k\rangle\equiv\sum_{\alpha=1}^{{\cal D}_k}
\tilde\psi_{k\alpha}\, ||k,\alpha;N-k,\alpha\rangle\rangle\ ;
\label{4-3}
\end{equation}
in this decomposition, the orthonormal vectors $\{||k,\alpha;N-k,\alpha\rangle\rangle\}$ form the
Schmidt basis, with ${\cal D}_k={\rm min}\{D_k, D_{N-k}\}$ ({\it cf.} (\ref{2-6})), while the Schmidt coefficients
$\tilde\psi_{k\alpha}$ are non negative real numbers, satisfying 
the normalization condition $\sum_k\sum_\alpha (\tilde\psi_{k\alpha})^2=1$.
In this new basis, one can easily compute the negativity of the density matrix $|\psi\rangle\langle\psi|$,
\begin{equation}
{\cal N}\big(|\psi\rangle\langle\psi|\big)=
{1\over2}\bigg[\bigg(\sum_{k=0}^N\sum_{\alpha=1}^{{\cal D}_k} \tilde\psi_{k\alpha}\bigg)^2-1\bigg]\ .
\label{4-4}
\end{equation}
Clearly, the negativity increases monotonically with the sum
$\sum_k\sum_\alpha \tilde\psi_{k\alpha}$, that therefore needs to be maximized
under the constraint $\sum_k\sum_\alpha (\tilde\psi_{k\alpha})^2=1$.
One easily shows that the maximum is obtained when all coefficients $\tilde\psi_{k\alpha}$ are equal and
constant,
\begin{equation}
\tilde\psi_{k\alpha}={1\over\sqrt{\cal D}}\ ,\qquad {\cal D}=\sum_{k=0}^N {\cal D}_k\ ,
\label{4-5}
\end{equation}
so that, ${\cal N}\big(|\psi\rangle\langle\psi|\big)=({\cal D}-1)/2$. Further, all Schmidt coefficients
need to be non vanishing in order to get this maximum value for $\cal N$; indeed, for any state
$|\psi'\rangle$ with Schimdt number ${\cal D}'< {\cal D}$, one has:
${\cal N}\big(|\psi'\rangle\langle\psi'|\big)<({\cal D}-1)/2$. \hfill $\Box$
\medskip

\noindent
In analogy with the standard case, states for which the negativity reaches the maximum value
$({\cal D}-1)/2$ can be called ``maximally entangled states''.

\medskip

\noindent
{\bf Remark 5:} {\sl i)} Given a fixed bipartition $(m, M-m)$, let us consider such
a maximally entangled state $\rho=|\psi\rangle\langle\psi|$. By tracing over the degrees of freedom
of the modes pertaining to the second partition, one obtains the reduced density matrix 
$\rho^{(1)}$ describing the first $m$ modes only:
\begin{equation}
\rho^{(1)}={1\over {\cal D}} \sum_{k=0}^N\sum_{\alpha=1}^{{\cal D}_k}
||k,\alpha\rangle\rangle\, \langle\langle k,\alpha||\ ;
\label{4-6}
\end{equation}
similarly, by tracing over the first partition, one obtains the reduced density matrix $\rho^{(2)}$
describing the second $N-k$ modes. 
In the case of distinguishable particles, either $\rho^{(1)}$ or $\rho^{(2)}$ 
is proportional to the identity matrix; this is no longer true here:
in fact, given a block with $k$ fixed, this happens only when $D_k\leq D_{N-k}$, {\it i.e.} when 
${\cal D}_k=D_k$.\hfill\break
{\sl ii)} On the other hand, given the expression (\ref{4-6}), one easily computes its von Neumann entropy,
obtaining: $S(\rho^{(1)})=-{\rm Tr}\big[\rho^{(1)}{\rm ln}\rho^{(1)}\big]={\rm ln}\,{\cal D}$,
as in the standard case.\hfill\break
{\sl iii)} Similarly, the purity of $\rho^{(1)}$ is given by: ${\rm Tr}\big[\big(\rho^{(1)}\big)^2\big]=1/{\cal D}$.
This result and that of the previous remark can equally be taken as alternative, equivalent definitions
of the notion of maximally entangled states.\hfill $\Box$
\medskip

Let us now come to the analysis of the space $\cal S$ of separable states
as determined by the generic bipartition $(m, M-m)$.
Among them, the totally mixed state $\rho_{\rm mix}$, proportional to the unit matrix, 
stands out because of its peculiar properties. In fact, recall that in the case
of distinguishable particles, $\rho_{\rm mix}$ always lays in the interior of $\cal S$
\cite{Zyczkowski,Bengtsson}; instead, now one finds:
\medskip

\noindent
{\bf Proposition 5.} {\sl Given any bipartition $(m, M-m)$ and an associated
separable basis made of the Fock vectors $|k, \sigma;N-k, \sigma'\rangle$,
the totally mixed state,
\begin{equation}
\rho_{\rm mix}={1\over D}\sum_{k=0}^N\ \sum_{\sigma,\sigma'}\
| k, \sigma; N-k, \sigma'\rangle \langle k, \sigma; N-k, \sigma' |\ ,
\label{4-7}
\end{equation}
lays on the border of the space $\cal S$ of separable states.}
\medskip

\noindent
{\sl Proof.} Let us take a state $\rho_{\rm ent}$ which can not be written in block diagonal form
as in (\ref{2-11}), (\ref{2-12}); by {\sl Proposition 1}, it is entangled.
Then, consider the combination
\begin{equation}
\rho_\epsilon={1\over 1+\epsilon}\big(\rho_{\rm mix}+\epsilon\, \rho_{\rm ent}\big)\ , \qquad \epsilon>0\ .
\label{4-8}
\end{equation}
For any $\epsilon$, this combination
will never be separable, since it is not block diagonal.
In other terms, in the vicinity of $\rho_{\rm mix}$, one always
find entangled states, so that it must lay on the border of $\cal S$.\hfill $\Box$
\medskip

\noindent
{\bf Remark 6:} {\sl i)} Note that similar considerations apply to all separable states:
there always exist small perturbations of separable, necessarily block diagonal, states that
make them not block diagonal, hence entangled. Instead, in the case of distinguishable
particles, almost all separable states remain separable under sufficiently small
arbitrary perturbations \cite{Bengtsson,Bandyopadhyay}.\hfill\break
{\sl ii)} Further, using analogous steps, one can show
that a Werner-like state, $\rho_W=p\, |\psi\rangle\langle\psi| + (1-p)\, \rho_{\rm mix}$,
$0\leq p\leq 1$, with $|\psi\rangle$ a maximally entangled state, 
is entangled for any nonvanishing value
of the parameter $p$, while for distinguishable $d$-level particles, this happens only when
$1/(d+1)<p\leq 1$ \cite{Werner1}.\hfill $\Box$
\medskip

The result of {\sl Proposition 5} is most strikingly illustrated by considering 
a system of $N$ bosons that can occupy
two modes ($M=2$), each of which forming a partition. In the previous section, we have seen
that a generic state 
$\rho$ for this system can be written in terms the Fock basis as in (\ref{3-17}); further,
it results separable if and only the density matrix $\rho$ is diagonal in this basis. 
Instead, $\rho_\epsilon$ given in (\ref{4-8})
will develop non-diagonal entries as long as $\epsilon$ starts to be non vanishing.

As mentioned before, in this case the set of $N+1$ vectors $\{|k;N-k\rangle\}$ 
constitutes the only basis made of separable pure states \cite{Benatti3}. 
Therefore, the decomposition of a generic separable state $\rho$ in terms
of projections on separable states turns out to be unique.
As a consequence, the set $\cal S$ of separable states is a sub-variety
of the convex space $\cal M$ of all states that turns out to be a simplex, 
whose vertices are given precisely by these projections.

The space of states of $N$ bosons confined in two modes
is then much more geometrically structured than in the case
of systems made of distinguishable particles.
Indeed, given a complete set of
observables $\{ {\cal O}_i \}$, one can decompose any density matrix $\rho$ as:
\begin{equation}
\rho=\sum_i\, \rho_i\ \Bigg( { \sqrt{\rho}\, {\cal O}_i\, \sqrt{\rho} \over
{\rm Tr}\big[{\cal O}_i\rho\big] }\Bigg)\ ,\qquad
\rho_i\equiv{\rm Tr}\big[{\cal O}_i\rho\big]\ ,\quad {\cal O}_i>0,\quad \sum_i {\cal O}_i={1};
\label{4-9}
\end{equation}
this decomposition is over pure states, whenever the operators ${\cal O}_i$ are chosen to be projectors.
Therefore, in general, there are infinite ways of expressing a density matrix
as a convex combination of projectors, even when these projectors are made of separable states.
As seen, this conclusion no longer holds for systems of identical particles.

The totally mixed state $\rho_{\rm mix}$ has also another interesting property:%
\footnote{For systems of distinguishable particles, the problem of finding so-called
``absolutely separable states'', {\it i.e.} states that are separable in any choice of
tensor product structure, has been discussed in \cite{Kus,Bengtsson}.}

\medskip
\noindent
{\bf Proposition 6.} {\sl The totally mixed state is the only
state that remains separable for any choice of bipartition.}

\medskip

\noindent
{\sl Proof.} First, let us again consider the
simplest case $M=2$. Any Bogolubov transformation maps the set of creation and annihilation
operators $a_i^\dagger$ and $a_i$, $i=1,2$ into new ones $b_i^\dag$, $b_i$, $i=1,2$;
a simple example is given by
\begin{equation}
b_1={a_1+a_2\over\sqrt{2}}\ ,\qquad
b_2={a_1-a_2\over\sqrt{2}}\ ,
\label{4-10}
\end{equation}
and their hermitian conjugates. The operators $b_i^\dag$, $b_i$
define a new bipartition $({\cal B}_1, {\cal B}_2)$ of the full algebra 
$\cal B$ of bounded operators, distinct from original one $({\cal A}_1, {\cal A}_2)$
generated by $a_i^\dag$, $a_i$. States (and operators as well) that are local
in one bipartition might turn out to be non-local in the other.
For instance, the Fock states $\{|k,N-k\rangle\}$ result entangled with respect
to the new bipartition defined by the transformation (\ref{4-10}); indeed, one finds:
\begin{equation}
|k,N-k\rangle={1\over 2^{N/2}}{1\over\sqrt{k!(N-k)!}}\sum_{r=0}^k\sum_{s=0}^{N-k}
{k\choose r}{N-k\choose s}(-1)^{N-k-s} \big(b_1^\dag\big)^{r+s}\, \big(b_2^\dag\big)^{N-r-s}\, |0\rangle\ ,
\label{4-11}
\end{equation}
so that $|k,N-k\rangle$ is a combination of $({\cal B}_1, {\cal B}_2)$-separable states. 
A similar conclusion applies to the mixed states (\ref{3-17}): in general, any
separable state $\rho_{\rm sep}=\sum_k \rho_k\, |k;N-k\rangle\langle k; N-k|$
is mapped by a Bogolubov transformation into a non-diagonal
density matrix, and therefore into an entangled state. 
In fact, one can always find a unitary transformation $U$ that maps any
diagonal matrix $\rho$ into a non-diagonal one $U\rho\, U^\dagger$. In particular,
when $\rho$ is not degenerate, the transformed matrix $U\rho\, U^\dagger$ results
diagonal only if the operator $U$ is itself a diagonal matrix; 
in this case however the corresponding Bogolubov transformation 
results trivial and does not define a new partition.
The only density matrix that
remains invariant under all unitary transformations
is the one proportional to the unit matrix, {\it i.e.}
the totally mixed one. This conclusion can easily be extended to the
multimode case: given any separable state in a given $(m, M-m)$ bipartition,
$\rho_{\rm sep}=\sum_{k=0}^N\ \sum_{\sigma,\sigma'}\ \rho_{k\sigma\sigma'}
| k, \sigma; N-k, \sigma'\rangle \langle k, \sigma; N-k, \sigma' |$,
one can always construct a Bogolubov transformation that maps it into
an entangled one: it is sufficient to apply the above considerations
to any couple of modes belonging to separate partitions. Therefore, also in this more general
setting, only the state $\rho_{\rm mix}$
in (\ref{4-7}) is left invariant by all Bogolubov transformations.\hfill $\Box$
\medskip

Thanks to these results, the global geometry of the space of $N$-boson states
starts to emerge more clearly. Again the two-mode case is easier to describe.
We have seen that by fixing a bipartition one selects the sets $\cal S$ of separable states;
these form a sub-variety of the convex space $\cal M$ of all states, forming
a $(N+1)$-dimensional simplex, with the projectors over Fock states as generators. 
Changing the bipartition through a Bogolubov transformation produces a new
simplex, having only one point in common with the starting one,
the state $\rho_{\rm mix}$. The geometry of the space of two-mode $N$-boson states
has therefore a sort of star-like topology, with the various simplexes sharing just one point,
the totally mixed state.

The case of $M$ modes is more complex. For a fixed bipartition,
the space of separable states is a sub-space of the convex space of all
states which is not any more strictly a simplex:
the decomposition of generic separable state $\rho$ is no longer unique,
since the Fock states in (\ref{2-6}) are no longer the only separable pure states:
for each $k$, reshufflings over the indices $\sigma$, $\sigma'$
are still allowed. Nevertheless, also in this case the global state space
presents a sort of star-like topology: only one point is shared by all separable bipartition sub-spaces,
the totally mixed state $\rho_{\rm mix}$.

\section{Outlook}

In many-body systems composed by a fixed number of identical particles,
the associated Hilbert space no longer exhibits the familiar particle tensor product structure.
The usual notions of separability and entanglement
based on such structure is therefore inapplicable: a generalized 
definition of separability is needed and can be given in terms
of commuting algebras of observables instead of focusing on
the particle tensor decomposition of states. 
The selection of these algebras is largely arbitrary, making it apparent
that in systems of identical particles entanglement
is not an absolute notion: it is strictly bounded 
to specific choices of commuting sets of system observables.

Using these generalized definitions, we have studied bipartite entanglement in systems composed
by $N$ identical bosons that can occupy $M$ different modes.
More specifically, we have analyzed to what extent entangled states
result robust against mixing with another state, either separable or entangled.
We have found that in general, the entanglement contained in bosonic states is much more
robust than the one found in systems of distinguishable particles. 
This result has been obtained by analyzing the
so-called {\sl robustness} and {\sl generalized robustness} of entanglement, of which explicit
expressions and upper bounds have been respectively given.
A quite general characterization of the geometry of the space of bosonic states has also been
obtained: this space exhibits a star-like structure composed by intersecting subspaces
each of one determined by a given bipartition
through the subset of separable states.
All these separable subspaces share one and only one point, the totally mixed state,
hence the star-shape topology.

As a final remark, notice that all above results can be generalized to the case
of systems where the total number of particles is not fixed,
but commutes with all physical observables ({\it i.e.} we are in
presence of a superselection rule \cite{Bartlett}).
In such a situation, a general density matrix $\rho$ can be written as an incoherent mixture 
of states $\rho_N$ with fixed number $N$ of
particles, having the general form (\ref{2-9}); explicitly:
\begin{equation}
\rho=\sum_N \lambda_N \rho_N \ ,\qquad \lambda_N\geq 0\ ,\qquad \sum_N \lambda_N=1\ .
\label{5-1}
\end{equation}
The state $\rho$ is thus a convex combination of matrices $\rho_N$ having support
on orthogonal spaces. Arguments similar to those used in proving {\sl Proposition 2}
(and {\sl Proposition 3}) allow us to conclude that both the robustness and the
generalized robustness of entanglement of the state $\rho$ in (\ref{5-1}) 
is the weighted average of the robustness of the components $\rho_N$, {\it i.e.}
$R(\rho)=\sum_N \lambda_N\, R(\rho_N)$ for both cases. The problem of computing the
robustness of incoherent particle number mixtures (\ref{5-1}) is then reduced 
to that of determining the robustness of the
corresponding components at fixed particle number, for which the considerations
and results discussed in the previous sections apply.


\end{document}